\documentclass[fleqn,twoside]{article}
\usepackage{espcrc2,cite}
\input{amssym.def}

\usepackage{epsf}
\usepackage[figuresright]{rotating}

\def\la{\langle}
\def\ra{\rangle}

\def\l{\left}
\def\r{\right}
\def\nn{\nonumber}

\def\ord#1{\mathcal{O}\l(#1\r)}

\def\cH{\mathcal{H}}
\def\cO{\mathcal{O}}
\def\cL{\mathcal{L}}

\def\re{\mathrm{Re}\;}

\def\beq{\begin{equation}}
\def\eeq{\end{equation}}
\def\bea{\begin{eqnarray}}
\def\eea{\end{eqnarray}}

\def\figcaption#1{\caption{\small\sl #1}}

\def\eq#1{Eq.\ {(\ref{#1})}}
\def\tab#1{Table {\ref{#1}}}
\def\sec#1{Section {\ref{#1}}}
\def\fig#1{Figure {\ref{#1}}}

\def\mev{\mbox{MeV}}
\def\gev{\mbox{GeV}}
\def\fm{\mbox{fm}}
\def\msbar{\overline{\mbox{MS}}}

\def\ln{\mathrm{ln}}

\title{Light Hadron Weak Matrix Elements
\vspace{-3.5cm}\begin{flushright}\small CPT-2000/P.4091
\\ LAPTH-Conf-821/00\end{flushright}
\vspace{2.2cm}}

\author{Laurent Lellouch\address{Centre de Physique Th\'eorique, Case 907, 
CNRS Luminy, F-13288 Marseille Cedex 9, France}%
}
       
\begin{document}

\begin{abstract}
I review this year's developments in the study of weak matrix elements
of light hadrons on the lattice, with emphasis on $K^0{-}\bar K^0$ mixing
and $K\to\pi\pi$ decays.

\end{abstract}

\maketitle

\section{Introduction}

Weak processes involving light hadrons in general, and kaons in
particular, have contributed significantly to our understanding of
fundamental interactions over the years.  Most recently, the
measurement of a non-vanishing $\re\epsilon'/\epsilon$ in $K\to\pi\pi$
decays by KTeV \cite{Alavi-Harati:1999xp} and NA48 \cite{Fanti:1999nm}
has provided unambiguous evidence for direct CP violation.

\begin{figure}[t]
\epsfxsize=8.5cm\epsffile{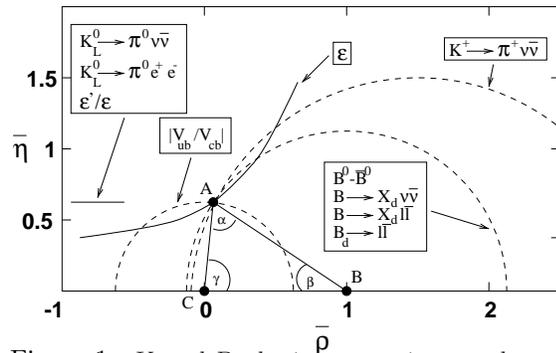}
\vspace{-1.2cm} \figcaption{\label{fig:triangle}\small\it $K$
and $B$ physics constraints on the unitarity triangle in a world
without errors (from \cite{Buras:1998ra}).}
\vspace{-0.7cm}
\end{figure}

In constraining the Standard Model (SM), the physics of kaons is
complementary to that of $B$ mesons. This is clearly visible in
\fig{fig:triangle}, where the types of constraints imposed by weak
processes involving either of these particles are displayed. Kaons
provide an important constraint on the summit of the unitarity
triangle through the measurement of $\epsilon$, the parameter which
quantifies indirect CP violation in $K\to\pi\pi$ decays. This
constraint requires a description of non-perturbative effects in
$K^0{-}\bar K^0$ mixing, parametrized by $B_K$. Lattice results for
this quantity are commonly used in unitarity triangle fits. There are
two new results this year for $B_K$, obtained with domain-wall
fermions, by the CP-PACS \cite{yusuke} and RBC \cite{tom}
collaborations.

Rare kaon decays and other FCNC kaon processes also provide good
probes of physics beyond the SM. Here too the lattice is
contributing. Donini {\it et al.}\ have computed SUSY, $\Delta S=2$
matrix elements \cite{Donini:1999nn} and the SPQcdR collaboration have
performed the first lattice determination of the electromagnetic
operator matrix element, $\la\pi^0|Q_\gamma^+|K^0\ra$, which
contributes to enhance the CP violating component of the $K_L\to\pi^0
e^+e^-$ amplitude in SUSY extensions of the SM
\cite{Becirevic:2000zi,guido}.

The need for a non-perturbative technique such as lattice QCD, to
study the processes discussed above is clear: the simple free-quark
picture of weak interactions is severely modified by the
non-perturbative effects of the strong interaction. One of the most
extreme examples of this is, of course, the $\Delta I=1/2$ rule in
$K\to\pi\pi$ decays: decays in which isospin changes by $\Delta I=1/2$
are greatly enhanced over those in which the change is $\Delta
I=3/2$. Using the operator product expansion (OPE) to separate short
and long distance contributions, one finds that most of the
enhancement, in a QCD based explanation, must come from
non-perturbative QCD effects in the matrix elements, $\la
(\pi\pi)_{I=0,2}|\cH_{\Delta S=1}|K^0\ra$, of the effective, $\Delta
S=1$ weak hamiltonian, $\cH_{\Delta S=1}$
\cite{Gaillard:1974nj,Altarelli:1974ex}.

The verification of this $\Delta I=1/2$ enhancement from first
principles, as well as the calculation of $\epsilon'/\epsilon$ and
more generally the study of non-leptonic weak decays are some of
the big challenges still facing lattice phenomenology. Though the problems
these processes pose are many, much has been learned over the years:

\medskip

1) The direct study of $K\to\pi\pi$ decays requires one to consider
four-point functions. These correlation functions are statistically
much more noisy than the two- and three-point functions encountered in
most phenomenological studies undertaken on the lattice. To reduce the
noise, one can consider transitions where all three particles are at
rest or even study $K\to\pi$ (and $K\to 0$) transitions. In both
cases, one uses chiral perturbation theory ($\chi$PT) to relate the
quantities computed on the lattice to the physical matrix elements.

\medskip

2) The renormalization of the dimension-six, four-quark
operators which appear in the effective weak hamiltonian is difficult
on the lattice:

$\bullet$ There is mixing with other dimension-six operators. Some of
this mixing is the same as in the continuum. But with fermion
formulations which break chiral symmetry explicitly, such as Wilson
fermions, there is additional mixing with wrong chirality
operators. This problem, however, is well understood and relatively
well controlled.

$\bullet$ There is also mixing with lower-dimension operators. This
gives rise to power divergences, proportional to inverse powers of
$a$, where $a$ is the lattice spacing. The subtraction of these
divergences is numerically very demanding. One must also be very
careful.  As pointed out some years ago by Martinelli, perturbation
theory is likely to fail for such subtractions. Indeed, the
coefficients of these divergences may have non-perturbative
contributions of the form
$e^{-\int^{g(a)}dg'/\beta(g')}=a\Lambda_{QCD}$, which are formally
smaller than any term in a perturbative expansion. However, when
enhanced by a power divergence, they can give a contribution which is
of the same size as the physical quantity that is being calculated. A
second concern is the definition of power subtracted operators in the
presence of discretization errors. Again, a discretization error of,
say, $\ord{a\Lambda_{QCD}}$, on a linearly-divergent quantity with
mass-dimension one may combine with this divergence and give a finite
contribution of $\ord{\Lambda_{QCD}}$.

Many tools have been developed over the years to address the problem
of renormalization (for details, I refer you to Stefan Sint's
contribution):

$\bullet$ Lattice perturbation theory, performed in terms of a
continuum-like, renormalized coupling constant can, in principle, be
used to subtract mixings with same-dimension operators.

$\bullet$ CPS symmetry \cite{Bernard:1985wf,Bernard:1988pr}, which is
a discrete symmetry of many four-quark weak operators, is a powerful
tool for classifying the mixings these operators may have. A CPS
transformation is a CP transformation, followed by a switching
transformation which changes $s$ into $d$ quarks and vice versa. This
symmetry is of course broken, but only softly by terms proportional
to powers of $(m_s-m_d)$. There are also generalizations of CPS in
which different pairs of quarks are swapped. I will generically refer
to all of these transformations as CPS.

$\bullet$ Chiral symmetry, in the form of $\chi$PT or of the sytematic
exploitation of chiral Ward identities can be used to guide, for
instance, the subtraction of power divergences in $K\to\pi$ amplitudes
for chirally symmetric lattice-fermion formulations
\cite{Bernard:1985wf} or the construction of effective weak
hamiltonians in the case where the lattice fermions break chiral
symmetry explicitly \cite{Bochicchio:1985xa,Maiani:1986rr}.

$\bullet$ Chiral symmetry is not sufficient to determine the
logarthmically divergent renormalizations required to make certain
quark bilinears or quadrilinears finite. For these subtractions, a
non-perturbative renormalization (NPR) technique was devised
\cite{Martinelli:1995ty}. It is performed with quark states and makes
use of a regularization-independent (RI) scheme. It requires gauge
fixing and the existence of a window $\Lambda_{QCD}\ll\mu\ll \pi/a$,
where $\mu$ is the renormalization scale. $\mu^2$ is usually taken to
be the momentum squared of the quarks. $\Lambda_{QCD}\ll\mu$ is
necessary to guarantee that the renormalization constants obtained are
independent of the states used to calculate them. $\mu\ll \pi/a$ keeps
discretization errors in check. This technique can also be used to
compute the coefficient required to subtract the mixing of
wrong-chirality operators in calculations involving four-quark
operators.

$\bullet$ Another NPR technique has been developed, which makes use of
a finite-volume, Schr\"odinger functional (SF) scheme
\cite{Jansen:1996ck}. It is associated with a non-perturbative
renormalization group scaling and overcomes the problems encountered
with the NPR techniques of \cite{Martinelli:1995ty}. It is gauge
invariant, requires no window and is mass independent. It has not yet
been applied to four-quark operators.

$\bullet$ Finally, to circumvent the mixing problem completely, the
authors of \cite{Dawson:1998ic} have suggested calculating the
hadronic matrix element of the $T$-product of weak currents,
$T\l\{J^W_\mu(x)J^{W\dagger}_\mu(0)\r\}$ for $a\ll |x|\ll
1/\Lambda_{QCD}$ on the lattice. The results would then be fitted to
the continuum OPE, giving directly the matrix elements of the
four-quark operators in the desired continuum scheme. Only the
renormalization of bilinears would be necessary.  The drawback is the
need for very small lattice spacings which makes this approach costly
numerically.

\medskip 

While much is already known, contributions are still being made to the
problem of renormalization and mixing. Two new suggestions were
proposed this year for obtaining $B_K$ with Wilson fermions but
without the subtraction of wrong-chirality contributions usually
required \cite{sintring,Becirevic:2000cy}. The one-loop mixing of
three- and four-quark operators with same-dimension operators has been
analyzed in the domain-wall formalism
\cite{Aoki:1999ky,Aoki:2000ee}. The renormalization of quark bilinears
and four-quark operators, for domain-wall fermions, has been performed
non-perturbatively in the RI scheme \cite{Dawson:2000yx}. The same
renormalization, for operators which have no power-divergences, has
been performed at one loop for Neuberger fermions
\cite{Alexandrou:2000kj,Capitani:2000da}.

\medskip

3) A third problem is that present day lattices are only a few fermi
long. On such lattices, it is not possible to separate the hadrons in
a multiparticle state into asymptotic states.

\medskip

4) A fourth problem is that the lattice method ``only'' provides
approximate, euclidean correlation functions. This forces us to
confront the ``Maiani-Testa theorem'' \cite{Maiani:1990ca} which, in
one of its guises, states that euclidean correlation functions which
describe processes involving two or more final-state hadrons in the
center of mass frame, yield amplitudes in which these particles are at
rest. Thus, an euclidean correlation function chosen to describe
$K\to\pi\pi$ decays will yield an amplitude for
$K(\vec{0})\to\pi(\vec{0})\pi(\vec{0})$, which clearly has the wrong
kinematics, unless $m_K=2m_\pi$. These statements will be made more
precise below.

\medskip

Both problems 3) and 4) can be addressed with $\chi$PT. For the first,
$\chi$PT enables estimates of leading finite-volume corrections.  It
helps with the second problem in that it provides a means of
extrapolating results to the correct physical point. In its quenched
or partially quenched versions, it further gives a handle on
quenching errors. There are new results in $\chi$PT this year which
are directly relevant for lattice studies of weak decays of kaons  and
which I briefly review in \sec{sec:chipt}.

While $\chi$PT seeks to correct finite-volume effects and problems
resulting from the ``Maiani-Testa theorem'', a new approach to
non-leptonic weak decays was proposed this year. It uses the fact
that lattice volumes are finite to circumvent the ``Maiani-Testa
theorem'' and yield directly $K\to\pi\pi$ matrix elements from
euclidean correlation functions \cite{Lellouch:2000pv}. Similar issues
are currently being investigated by Lin {\it et al.}\
\cite{linetal,Testa:2000ce}. 

\medskip

Before closing this rather lenghty introduction, I would like to say a
few words about domain-wall (and Neuberger) fermions, since they are
beginning to play a prominent r\^ole in the study of light-hadron weak
matrix elements. For details I refer you to Pavlos Vranas' talk
\cite{vranas}. These two recent formulations of lattice fermions have
a significant advantage over the more traditional Wilson and
Kogut-Susskind formulations. They have a full chiral-flavor symmetry
at finite lattice spacing. There is, of course, a price to pay. For
domain-wall fermions it is a fifth dimension with $N_5\to\infty$
sites; for Neuberger fermions it is the inverse square root of a large
matrix. The question then becomes: how small an $N_5$ or poor an
approximation to the inverse square root can one take and still have
sufficient chiral symmetry? Let me concentrate on domain-wall
fermions, which are currently more commonly used for weak-interaction
phenomenology.  The assessment is somewhat mixed. The mathematical
convergence in $N_5$, which is exponential asymptotically, can be
rather slow for the couplings and volumes commonly used, though
improvements can be made
\cite{Edwards:2000qv,Hernandez:2000iw}. However, it appears that in
practice, for currently used simulation parameters, the chiral
symmetry achieved may be sufficient for studying the physics of quarks
whose masses are around that of the strange
\cite{Blum:2000kn,AliKhan:2000iv,Jung:2000fh}. The criterion used in
these studies is the residual mass, $m_{res}$, which essentially
measures the amount the bare quark mass has to be shifted away from
zero in order to have physically massless quarks.  At $\beta=6.0$ with
$N_5=16$ and on a $16^3\times 32$ lattice, the authors of
\cite{Blum:2000kn} find that $m_{res}(\msbar,2\,\gev)\simeq
4\mev$. Nevertheless, it should be kept in mind that the amount of
chiral symmetry required will depend on the quantity studied. Indeed,
the exponentially small chiral symmetry breaking can combine with
power divergences induced by this breaking to give effects which may
no longer be considered negligible. This problem is certainly relevant
for matrix elements of four-quark operators whose mixings with
lower-dimensional operators are often made significantly worse when
chiral symmetry is broken explicitly. This issue is currently being
addressed by the RBC collaboration \cite{bob}.

\medskip

Given the additional cost and potential difficulties of
chirally-improved fermions, one is entitled to ask where, in weak
matrix element calculations, is exact or nearly exact chiral symmetry
absolutely necessary?  This question is all the more justified that
one appears to be able to access the desired physics with Wilson or
Kogut-Susskind fermions in most cases. These calculations, however,
are sufficiently complex and difficult that it is not clear, a priori,
whether the overhead associated with chirally-improved fermions cannot
be offset, at least partially, by their improved chiral behavior, by
the fact that they are non-perturbatively $\ord{a}$-improved, etc. It
is thus important that all approaches be pursued.

\medskip

The rest of this review is organized as follows. In \sec{sec:chipt}, I
review two new studies in $\chi$PT which are relevant for lattice
calculations of weak decays of kaons. In \sec{sec:kkbar}, I discuss
$K^0{-}\bar K^0$ mixing. A brief description of the $\Delta S=1$
effective hamiltonian is given in \sec{sec:kpipigc}. Calculations of
$K\to\pi\pi$ matrix elements from $K\to\pi$ and $K\to 0$ amplitudes
are reviewed in \sec{sec:kpik0}, including two new ambitious studies
using domain-wall fermions. In \sec{sec:kpipi}, I discuss some of the
issues surrounding the calculation of physical $K\to\pi\pi$ matrix
elements from lattice $K\to\pi\pi$ amplitudes. A new approach to the
calculation of non-leptonic weak decays is presented in
\sec{sec:finitev}. \sec{sec:ccl} contains my conclusions.

\medskip
The focus of this talk is on recent lattice developments. I will
unfortunately not have the space to cover the results obtained by
other methods. Please see \cite{Bertolini:2000dy} and references
therein for descriptions of some of the other possible non-perturbative
approaches.

\section{Chiral perturbation theory results}
\label{sec:chipt}

There are at least two new studies in $\chi$PT this year which
are directly relevant for lattice calculations of weak decays of kaons.

In the first, results for the one-loop, $\ord{p^2}$ corrections to the
$K\to\pi$ and $K\to\pi\pi$ matrix elements of the electroweak penguin
operators $Q_7$ and $Q_8$ (see \eq{eq:ds1ops}), which belong to the
$(8,8)$ representation of $SU(3)_L\times SU(3)_R$, were calculated
\cite{Cirigliano:2000pv}. The authors find that the $K\to\pi\pi$
amplitudes, including $\ord{p^2}$ counterterms, can be obtained from a
study of the $m_K^2$, $m_\pi^2$ and $p_\pi\cdot p_K$ dependence of
$K\to\pi$ matrix elements.  They further find that chiral loops can
give rather significant ($+(27\pm 27)\%$) corrections to the
$\ord{p^0}$ results for $\la(\pi\pi)_{I=2}|Q_{7,8}|K\ra$. These
results are obtained in regular $\chi$PT. It would be interesting to
see how they are modified in quenched (q) or partially quenched (pq)
$\chi$PT.

In the second study, it is one-loop corrections to the matrix elements
of the $(8,1)$ and $(27,1)$ operators for $K\to\pi$ transitions at
rest with degenerate $s$ and $d$ quarks and for $K\to 0$ with $m_s\neq
m_d$, and the relation of these matrix elements to $K\to\pi\pi$
amplitudes, which are studied in pq$\chi$PT
\cite{Golterman:2000fw}. Not surprisingly, they find that $K\to\pi$
and $K\to 0$ amplitudes are not sufficient to determine all
$\ord{p^4}$ couplings required to obtain $K\to\pi\pi$ matrix elements
at this order. Furthermore, they observe that the chiral logarithms
are typically large and can depend strongly on $N_{sea}$. By using
their results to guide extrapolations of lattice results for these
amplitudes to the chiral limit, one can hope to obtain reliable
results for the $\ord{p^2}$ octet and twenty-seven-plet couplings,
which are interesting quantities in their own right and for which a
number of phenomenological estimates are available.

\section{$K^0{-}\bar K^0$ mixing}
\label{sec:kkbar}

The most general $\Delta S=2$ effective hamiltonian for $K^0{-}\bar
K^0$ mixing can be written in terms of the following operators:
\bea
\cO_1 &=&[\bar s_\alpha d_\alpha]_{V-A}[\bar s_\beta d_\beta]_{V-A}\nn\\
\cO_2 &=&[\bar s_\alpha d_\alpha]_{S-P}[\bar s_\beta d_\beta]_{S-P}\nn\\
\cO_3 &=&[\bar s_\alpha d_\beta]_{S-P}[\bar s_\beta d_\alpha]_{S-P}
\label{eq:ds2basis}\\
\cO_4 &=&[\bar s_\alpha d_\alpha]_{S-P}[\bar s_\beta d_\beta]_{S+P}\nn\\
\cO_5 &=&[\bar s_\alpha d_\beta]_{S-P}[\bar s_\beta d_\alpha]_{S+P}\nn
\ ,\eea
where $\alpha$ and $\beta$ are color indices.  In the SM, only $\cO_1$
contributes; in extensions, such as the MSSM, the other operators may
also be required.

These operators have positive and negative parity components, and only
the former contribute to $K^0{-}\bar K^0$ mixing. For instance,
\beq
\cO_1=\cO_{VV+AA}-2\cO_{VA}\ ,
\eeq
and $\cO_{VV+AA}$ determines the SM contributions to this mixing. As
is well known, the explicit chiral symmetry breaking present
in the Wilson formulation of fermions implies that $\cO_{VV+AA}$ will have
finite mixings with wrong chirality operators:
\beq
\cO_{VV+AA}(\mu)=Z_{VV+AA}
(a\mu,g^2)
\eeq
$$
\times\l\{\cO_{VV+AA}(a)+\sum_i\Delta_i(g^2)\cO_i\r\}\ ,
$$
with $i=VV-AA,\,SS+PP,\,SS-PP,\,TT$. This mixing is particularly bad
here because $\cO_{VV+AA}$ is subdominant in the chiral expansion.

\subsection{Two proposals for getting around mixing with wrong chirality
operators}
\label{sec:bkprops}

As shown by Bernard and collaborators many years ago, CPS symmetry
protects the parity odd components of the operators in \eq{eq:ds2basis} from
mixing with operators of wrong chirality \cite{Bernard:1988pr}. Thus,
\beq
\cO_{VA}(\mu)=Z_{VA}(a\mu,g^2)\cO_{VA}(a)
\ .
\eeq
So the basic idea behind these proposals is to relate the matrix
element of interest, $\la\bar K^0|\cO_{VV+AA}|K^0\ra$, to a correlation
function where the only four-quark operator is $\cO_{VA}$.

The basic ingredient is the chiral rotation \cite{sintring}
\beq
\l(\begin{array}{c}u\\d\end{array}\r)\to
\mathrm{exp}\l[i\alpha\gamma_5\tau^3/2\r]
\l(\begin{array}{c}u\\d\end{array}\r)
\label{eq:chirot}\ .
\eeq
It is implemented in two different ways.

In the first, one considers twisted-mass QCD \cite{Frezzotti:2000vv}:
\beq
\cL=\bar\psi\l(D_W+m_0+i\mu_0\gamma_5\tau^3\r)\psi
\eeq
$$
+\bar s\l(D_W+m_0^s\r)s
\ ,
$$
where $D_W$ is the usual Wilson Dirac operator, $\psi$ is the $(u,d)$
doublet and $m_0$, $\mu_0$ and $m_0^s$ are bare mass parameters. In
the continuum, this lagrangian would be equivalent to the usual,
three-flavor QCD lagrangian. On the lattice, however, because of the
Wilson term, it is not. The authors then suggest adjusting the
renormalized mass $m$ and twisted mass $\mu$ in such a way that the
angle of the rotation, $\alpha$, in \eq{eq:chirot} is $\pi/2$. Then,
the physical $\cO_{VV+AA}$ is $-2\cO_{VA}$ in the twisted theory,
which means that only a multiplicative renormalization is required.

The authors of \cite{Becirevic:2000cy} consider a Ward identity
associated with an infinitesimal version of the rotation of
\eq{eq:chirot}: 
$$
Z_P\,\la P(t_1)\cO_{VV+AA}(\mu;0) P(t_2)\ra= Z_{VA}(a\mu)
$$
$$
\times\bigl[2m\,Z_P
\sum_x\la\Pi(x) P(t_1)\cO_{VA}(a;0) P(t_2)\ra
$$
\beq
-Z_S\la S(t_1)\cO_{VA}(a;0) P(t_2)\ra 
\eeq
$$
-Z_S\la P(t_1)\cO_{VA}(a;0) S(t_2)\ra\bigr]+\ord{a}
\ ,
$$
with $P=\sum_{\vec{x}}\bar d\gamma_5s$, $S=\sum_{\vec{x}}\bar d s$,
$\Pi=\bar d\gamma_5d-\bar u\gamma_5u$ and $m =Z_A\la 0|\partial_\mu 
A_\mu^\pi|\pi^0\ra/2\la 0 |P^\pi|\pi^0\ra$.

Correlators containing the operator $S$ are exponentially suppressed
for $t_1,\,-t_2\to\infty$, since they are dominated by scalar instead
of pseudoscalar contributions. Thus, the correlator $\sum_x\la\Pi(x)
P(t_1)\cO_{VA}(0) P(t_2)\ra$ yields $\la\bar K^0|\cO_{VV+AA}(\mu)|K^0\ra$
without mixing.  Exploratory results for $B_K$ using this method were
presented at this conference \cite{guido}.

Both methods are generalizable to other matrix elements. Of course,
none of this is necessary with fermion formulations which have a
chiral symmetry, such as domain-wall, Neuberger or Kogut-Susskind
fermions.

\subsection{Matrix elements for $K^0{-}\bar K^0$ mixing beyond the Standard
Model}
\label{sec:ds2bsm}

Donini {\it et al.}\ have computed the matrix elements of all operators
of \eq{eq:ds2basis} between $K^0$ and $\bar K^0$ states
\cite{Donini:1999nn}. They work with a tree-level, $\ord{a}$-improved
Sheikholeslami-Wohlert (SW) action at $\beta=6.0$ and 6.2, in the
quenched approximation. Their strange and down quarks are
degenerate. They renormalize the matrix elements non-perturbatively in
the RI scheme, match them onto other schemes at one loop and run
them at two loops.

To quantify and subtract residual artifacts that might remain in
chiral behavior of the matrix element of $\cO_1$, they fit its 
dependence on mass and momenta to the form:
\beq
\frac{\la\bar K^0|\cO_1(\mu)|K^0\ra}{\frac{8}{3}f_K^2}=\alpha+\beta m_K^2
+\gamma p_K\cdot p_{\bar K}+\cdots
\ ,\label{eq:bksub}
\eeq
where the parameters $\alpha$ and $\beta$ are pure artifacts and the
dots include higher-order terms in the chiral expansion to which they
are not sensitive numerically.~\footnote{The fit they actually perform
corresponds to a rescaled version of \eq{eq:bksub}.} $B_K$ is
then simply $\gamma$.

In analogy with the definition of $B_K=\la\bar
K^0|\cO_1|K^0\ra/\frac{8}{3}f_K^2 m_K^2$, they suggest the following
normalization for the matrix elements of the other operators
($i=2,\cdots,5$):
\beq
\tilde B_i(\mu)=
\frac{\la\bar K^0|\cO_i(\mu)|K^0\ra}{f_K^2m_{K^*}^2}
\ ,\label{eq:ds2norm}
\eeq
instead of the usual, vacuum saturation approximation (VSA)
normalization, which is traditionally expressed in terms of the
quark-mass combination $[m_s(\mu)+m_d(\mu)]^2$. The problem with
giving the matrix elements in units of their VSA values is that they
are then used, in phenomenological applications, with uncorrelated
values of the quark masses, thus compounding the uncertainty on the
matrix elements with those on the poorly measured quark masses. This
problem obviously disappears with the normalization of
\eq{eq:ds2norm}.

The authors do observe some lattice spacing dependence in their
results, which never exceeds two statistical standard deviations. They
choose to average their $\beta=6.0$ and 6.2 results and keep the
largest statistical error. Their final results are summarized in
\tab{tab:doninids2}. It is worth noting that at $2\,\gev$ in the
$\msbar$-NDR scheme, the non-SM matrix
elements are typically 2 to 12 times larger than the matrix element of
$\cO_1$.
\begin{table}[t]
\caption{\label{tab:doninids2}\small\it Matrix elements of the $\Delta
S=2$ operators of \eq{eq:ds2basis} between $K^0$ and $\bar K^0$ states
as obtained in \cite{Donini:1999nn} in the RI scheme. Also given are
the corresponding renormalization group invariant (RGI) matrix
elements, as defined in \cite{Donini:1999nn}.}
\begin{center}
\begin{tabular}{ccc}
\hline\hline
$\la   \cO_i\ra$ & RI ($\mu=2\,\gev$) & RGI \\
\hline
$\la   \cO_1 \ra$ & $0.012(3)\,\gev^4$  & $0.017(4)\,\gev^4$ \\ 
$\la   \cO_2 \ra$ & $-0.077(10)\,\gev^4$ & $-0.050(7)\,\gev^4$\\ 
$\la   \cO_3 \ra$ & $0.024(3)\,\gev^4$   & $0.001(7)\,\gev^4$\\ 
$\la   \cO_4 \ra$ & $0.142(12)\,\gev^4$  & $0.068(6)\,\gev^4$\\ 
$\la   \cO_5 \ra$ & $0.034(5)\,\gev^4$   & $0.038(5)\,\gev^4$\\ 
\hline \hline
\end{tabular}
\end{center}
\vspace{-0.6cm}
\end{table}
The result $\la\cO_1\ra$ in \tab{tab:doninids2} corresponds to the 
following value for $B_K$:
\beq
B_K^{NDR}(2\,\gev)=0.68(21)
\ .
\eeq

\subsection{$B_K$ with domain-wall fermions}

As already pointed out above, chiral symmetry facilitates the
calculation of $B_K$. Domain-wall fermions are thus a good candidate
for such a calculation.  The first study of $B_K$ with domain-wall
fermions was actually performed some years ago \cite{Blum:1997mz}. Two
new quenched results were presented at this conference, by the CP-PACS
\cite{yusuke} and the RBC \cite{tom} collaborations. Both calculations
are performed with Shamir's variant of domain-wall fermions and
degenerate $s$ and $d$ quarks with masses approximatively ranging from 
$m_s/4$ to $m_s$. They differ, however, in the gauge action
employed. CP-PACS use an RG-improved action at two values of
$\beta=2.6$ and 2.9, corresponding to an inverse lattice spacing of
$1.81(4)\,\gev$ and $2.87(7)\,\gev$, respectively, as determined from
the $\rho$-meson mass. RBC used a standard Wilson plaquette action,
with $\beta=6.0$, corresponding to an inverse lattice spacing of
roughly $2\,\gev$. Another important difference in the calculations
is that RBC renormalize the matrix element non-perturbatively
\cite{Dawson:2000yx}, using the techniques of \cite{Martinelli:1995ty},
while CP-PACS match their results onto the continuum perturbatively,
at one loop \cite{Aoki:1999ky}.

CP-PACS perform a rather extensive study of the dependence of their
results on fifth-dimensional and spatial sizes and on cutoff. At $\beta=2.6$,
they work with the following four lattices: $(16{-}24{-}32)^3\times
40\times 16$ and $24^3\times 40\times 32$. At $\beta=2.9$, they have
the two lattices $(24{-}32)^3\times 60\times 16$. To investigate the
chiral properties of their $\Delta S=2$ matrix element, they consider
the ratio
\beq
B_P=
\frac{\la\bar K^0|\cO_1|K^0\ra}{\la \bar K^0|\bar s\gamma_5 d|0\ra
\la 0|\bar d\gamma_5 s|K^0\ra}
\ ,
\label{eq:bpdef}
\eeq
which should vanish, by chiral symmetry, in the limit $m_K\to 0$. They
study $B_P$ in the limit of vanishing quark mass, obtained by linear
extrapolation from finite quark mass. They find that extending the
fifth dimension by a factor of two, from 16 to 32 points, does not
reduce the deviation of $B_P$ from zero which they observe. Going to
smaller lattice spacing does not reduce the effect either whereas
increasing the volume does. The effect is of order -10 to -20\% in
units of $B_P$ at $m_K$ on their $(32{-}24{-}16)^3\times 40\times 16$
lattices.~\footnote{It will be slightly larger for $B_P$ extrapolated
to $m_q=-m_{res}$.} They conclude that the deviation of $\la\bar
K^0|\cO_1|K^0\ra$ from zero at vanishing quark mass must be a
finite-volume effect. This is supported by the fact that the volume
dependence of $B_P$ at finite mass increases rapidly as this mass is
reduced. At $m_K$, the reduction in $B_P$ in going from the largest to the
smallest volume is small, less than 2\%. At $am_q=0.01$, it is
approximatively 15\%.

\begin{figure}[t]
\centerline{\epsfxsize=7.5cm\epsffile{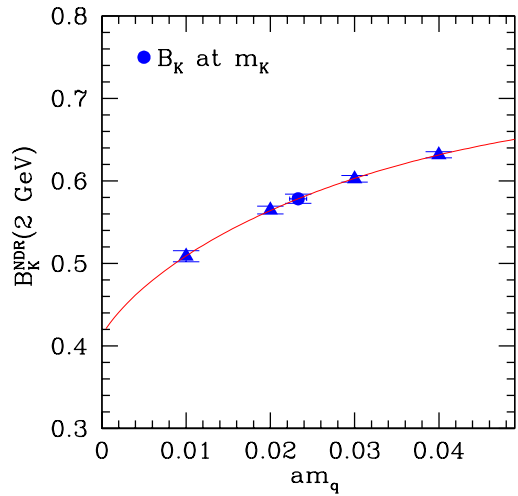}}
\vspace{-1.5cm}
\caption{\label{fig:bkmassdep}\small\it Dependence of
$B_K^{NDR}(2\,\gev)$ on bare quark mass, from CP-PACS
\cite{yusuke}.}
\vspace{-0.6cm}
\end{figure}

They then study the mass-dependence of $B_K$, and interpolate to the kaon
mass using the following $\chi$PT-inspired functional form
\beq
B_K=B(1-3c\times am_q\,\ln(am_q)+b\times am_q)
\ ,
\eeq
as shown in \fig{fig:bkmassdep} for their $24^3\times 40\times 16$
lattice. The physical point is obtained at half the strange quark
mass, as estimated from the experimental value of $m_K/m_\rho$. The
value of $B_K$ extrapolated to the chiral limit appears to be
approximatively 30\% smaller than the value at $m_K$, which would help
reconcile lattice results with the chiral-limit result for $B_K$
obtained recently in a large-$N_c$ approximation to QCD
\cite{Peris:2000sw}. However, it is important to note that
finite-volume effects of the kind described above could significantly
distort the chiral extrapolation of $B_K$.  If so, an extrapolation to
infinite volume would be necessary to determine $B_K$ reliably in the
chiral limit.

Finally, they study the dependence of $B_K$ at $m_K$ on
lattice-spacing, spatial volume and fifth dimensional size. They find
that these dependences are small. They obtain the following result for
$B_K$
\beq
B_K^{NDR}(2\,\gev)=0.575(6)
\label{eq:bkcppacs}
\ ,\eeq
by fitting, to a constant in lattice spacing, the results that they
obtain from their runs on a $24^3\times 40\times 16$ lattice at
$\beta=2.6$ and on a $32^3\times 60\times 16$ lattice at $\beta=2.9$.
Errors in \eq{eq:bkcppacs} are statistical only.

\medskip

From their fixed lattice spacing calculation on a $16^3\times 32\times
16$ lattice, the RBC collaboration obtain the following result for
$B_K$:
\beq
B_K^{NDR}(2\,\gev)=0.538(8)
\label{eq:bkrbc}
\ .\eeq
This result is lower than the CP-PACS result. This difference may be
due to the fact that RBC renormalize their results perturbatively
whereas CP-PACS do so perturbatively. The non-perturbative matching
coefficient used by RBC is 7\% smaller than the corresponding
mean-field improved, one-loop coefficient computed in
\cite{Aoki:1999ky}.

\subsection{$B_K$ summary}

\begin{figure}[t]
\centerline{\epsfxsize=7.5cm\epsffile{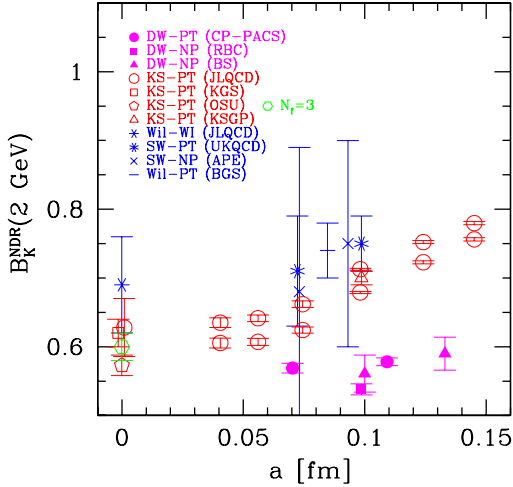}}
\vspace{-1.5cm}
\caption{\label{fig:bksummary}\small\it $B_K^{NDR}(2\,\gev)$ with
domain-wall (filled symbols), Kogut-Susskind (open symbols) and
Wilson/Sheikholeslami-Wohlert (stick symbols) fermions. All results
are quenched except for the $N_f=3$ result of OSU and all errors are
statistical. PT stands for perturbative matching, NP for
non-perturbative matching \`a la \cite{Martinelli:1995ty} and WI,
for matching using Ward identities. The references for the different
results are: CP-PACS \cite{yusuke}, RBC \cite{tom}, BS
\cite{Blum:1997mz} (with NPR, hep-lat/9909108), JLQCD (KS)
\cite{Aoki:1998nr}, KGS \cite{Kilcup:1998ye}, OSU and $N_f=3$
\cite{Kilcup:1997hp}, KSGP \cite{Kilcup:1990fq}, JLQCD (Wil)
\cite{Aoki:1999gw}, UKQCD \cite{Lellouch:1999sg}, APE
\cite{Donini:1999nn}, BGS \cite{Gupta:1997yt}.}
\vspace{-0.6cm}
\end{figure}

I summarize, in \fig{fig:bksummary}, the results for
$B_K^{NDR}(2\,\gev)$ obtained by different collaborations with
different fermion formulations. The size of the error bars on the
Wilson results is mostly a reflection of the subtractions which are
necessary to restore correct chiral behavior. The fact that different
actions give different results at fixed lattice spacing is a sign that
for some of the formulations at least, there is still some way to the
continuum limit.  In this limit, however, all formulations agree
within one and some standard deviations. The domain-wall results
exhibit rather good scaling. They are also systematically lower at
fixed lattice spacing. If this trend survives further scrutiny of the
continuum extrapolation and of systematic effects, we may have to
slightly revise our canonical estimate of $B_K$. However, for the
moment, because it involves the most extensive study of systematic
errors, the KS result of the JLQCD collaboration \cite{Aoki:1998nr}
should still be taken as the reference result. To this one must add
uncertainties due to quenching and the fact that the down and strange
quarks are degenerate in the calculation. Sharpe \cite{Sharpe:1998hh},
on the basis of q$\chi$PT and the preliminary ``$N_f=3$'' OSU results
\cite{Kilcup:1997hp}, suggests an enhancement factor of $1.05\pm 0.15$
to ``unquench'' quenched results for $B_K$. He advocates an additional
factor $1.05\pm 0.05$ to compensate for the effect of working with
degenerate quark masses. Here, we choose to keep his estimate of
errors, but not use his enhancement factors so as not to modify the
central value on the basis of information which is, for the most part,
not provided by lattice calculations.  Thus, we quote
\bea
&B_K^{NDR}(2\,\gev)=0.628(42)(99)& \\
&\to \hat B_K^{NLO}=0.86(6)(14)\ ,&\nn
\eea
where $\hat B_K^{NLO}$ is the two-loop RGI $B$-parameter obtained from
$B_K^{NDR}(2\,\gev)$ with $N_f=3$ and $\alpha_s(2\,\gev)=0.3$.
Of course, future studies should investigate systematically the effect of
including light flavors of dynamical quarks.

\section{$K\to\pi\pi$ decays: general considerations}
\label{sec:kpipigc}

With all massive modes, including charm, integrated out, the $\Delta
S=1$ effective hamiltonian is given by~\footnote{For a review see, 
for instance, \cite{Buras:1998ra}.}
\beq
\cH_{\Delta S=1} = \frac{G_F}{\sqrt{2}}V_{ud}V_{us}^*
\sum_{i=1}^{10}(z_i+\tau y_i)Q_i
\ ,
\eeq
where $z_i$ and $y_i$ are short-distance coefficients, $\tau
=-V_{ts}^*V_{td}/V_{us}^*V_{ud}$ and
\bea
Q_1 &=& (\bar s_{\alpha} u_{\beta})_{V-A}\;
(\bar u_{\beta} d_{\alpha})_{V-A}\nn\\
Q_2 &=& (\bar s u)_{V-A}\;(\bar u d)_{V-A}\nn\\
Q_{3,5} &=& (\bar s d)_{V-A}\sum_q(\bar qq)_{V\mp A}\nn\\
Q_{4,6} &=& (\bar s_{\alpha} d_{\beta})_{V-A}\sum_q(\bar q_{\beta} 
       q_{\alpha})_{V\mp A}\label{eq:ds1ops}\\
Q_{7,9} &=& {3\over 2}\;(\bar s d)_{V-A}\sum_qe_q\;(\bar qq)_{V\pm A}\nn\\
Q_{8,10} &=& {3\over2}\;(\bar s_{\alpha} d_{\beta})_{V-A}\sum_qe_q
        (\bar q_{\beta} q_{\alpha})_{V\pm A}\nn
\ ,
\eea
with $q=u,d,s$. It proves
useful to split the $\Delta S=1$ operators into a sum of operators
which transform under irreducible representations of the isospin
group: $Q_i=Q_i^{1/2}+Q_i^{3/2}$.

Integrating out charm is questionable since $m_c\sim 1.3\,\gev$.  Above
the charm threshold, one has the following effective hamiltonian:
\bea
\cH^c_{\Delta S=1} &=& \frac{G_F}{\sqrt{2}}V_{ud}V_{us}^*
\l[(1-\tau)\sum_{i=1}^{2}z_i(Q_i-Q_i^c)\r. \nn\\
& & \l. + \tau\sum_{i=1}^{10}(z_i+y_i)Q_i\r]
\ ,
\eea
where $Q_{1,2}^{c}=Q_{1,2}[u\to c]$, the sum over $q$ now runs over
$u,d,s,c$ and $y_i$ and $z_i$ are given $N_f=4$ values.

The r\^ole of the lattice is to compute:
\beq
\la Q_i\ra_I\equiv \la(\pi\pi)_I|Q_i|K\ra
\label{eq:ampdef}
\eeq
for the two isospin channels $I=0,2$ and $i=1,\cdots,10$, where
$Q_{1,2}$ in \eq{eq:ampdef} are to be understood as
$Q_{1,2}-Q_{1,2}^c$ in the case of an active charm.~\footnote{The
conventions adopted here for the normalization of isospin amplitudes etc.
are those of \cite{Buras:1998ra}.}

\section{$K\to\pi\pi$ from $K\to\pi$ and $K\to 0$}
\label{sec:kpik0}

The idea here is to compute three-point functions to obtain matrix
elements of $\Delta S=1$ operators between $K$ and $\pi$ states, where
the light quarks typically have masses around $m_s/2$, 
and then use $\chi$PT to relate them to the corresponding, physical
$K\to\pi\pi$ matrix elements. This last step is usually performed at
lowest non-trivial order. In the case of $\Delta I=1/2$ transitions,
there is an unphysical contribution to $K\to\pi$ matrix elements of
$(8,1)$ operators which does not appear in its physical $K\to\pi\pi$
counterpart. This contribution can be subtracted by considering $K\to
0$ transitions \cite{Bernard:1985wf}.

\subsection{Electroweak penguins with quenched Wilson fermions}

As part of their study of $\Delta S=2$ matrix elements, Donini {\it et
al.}\ \cite{Donini:1999nn} have computed the $\Delta I=3/2$,
electroweak-penguin matrix elements, $\la Q_{7,8}\ra_2$, which give
important contributions to $\re \epsilon'/\epsilon$. The parameters of
the calculation are the same as those presented in \sec{sec:ds2bsm}.

They use leading order $\chi$PT to get the $K\to\pi\pi$ matrix
elements in the chiral limit through
\beq
\la Q_{7,8}\ra_2=
\frac{\sqrt{3}m_\rho^2 f_\pi}{2}\lim_{m_q\to 0}\tilde B_{5,4}
\ ,\eeq
which they obtain by linear extrapolation of their $K^0-\bar K^0$
results calculated for degenerate quark masses, $m_s=m_d=m_q$. Here,
$f_\pi=131\,\mev$.  They find, in the chiral limit and at $2\,\gev$ in
the NDR-$\msbar$ scheme:~\footnote{It should be remembered that
the use of different normalizations for these matrix elements will
yield different values for the matrix elements in units of $\gev^3$.}
\beq 
\begin{array}{l}
\la Q_7(2\,\gev)\ra_2^{NDR}/(m_\rho^2 f_\pi)=1.7(3)\ ,\\
\la Q_8(2\,\gev)\ra_2^{NDR}/(m_\rho^2 f_\pi)=8.1(8)
\ .\end{array}
\label{eq:q78donini}
\eeq
These values are obtained by one-loop matching from their
non-perturbatively renormalized results in the RI scheme. This
matching is rather poorly behaved as it induces a 40\% change in $\la
Q_7\ra_2$ and a 27\% change in $\la Q_8\ra_2$. Since all other results
for these matrix elements are, to date, normalized by their vacuum
saturation value and this value is not provided, I will not attempt a
comparison with \eq{eq:q78donini}.

\subsection{Matrix elements of the $\Delta S=1$ operators from
quenched domain-wall QCD}
\label{sec:kpipifromdwf}

At this conference, the CP-PACS \cite{junichi} and RBC \cite{tom,bob}
collaborations presented preliminary results from calculations of the
matrix elements of all 10, $\Delta S=1$ operators. They use
domain-wall fermions and work in the quenched approximation. They are
also considering the case of active charm. Both collaborations work on
lattices of size $16^3\times 32\times 16$, with a domain-wall height
$M=1.8$ and cutoffs $a^{-1}\simeq 2\,\gev$. In addition to a Wilson
plaquette action at $\beta=6.0$ used by the two teams, CP-PACS also
perform the calculation with a RG-improved action at
$\beta=2.6$.~\footnote{For RBC, the gauge ensemble is the same as the
one used in their $B_K$ calculation. } In these calculations, a
pseudoscalar meson composed of degenerate quarks of bare mass
$am_q\sim 0.02$ would have a mass close to that of the physical
kaon. While CP-PACS renormalize their matrix elements at one loop
\cite{Aoki:2000ee}, RBC perform this renormalization
non-perturbatively \cite{Dawson:2000yx}.

\medskip

$\Delta I=3/2$ transitions involve no
``penguin'' contractions and therefore no mixing with lower
dimensional operators. And because of the approximate chiral symmetry
of domain-wall fermions, the structure of the renormalization is the
same as in the continuum, up to corrections which are exponentially
small in $N_5$.

$\Delta I=1/2$ transitions can receive contributions from lower
dimensional operators.  To leading order in the chiral expansion, the
matrix elements $\la \pi^+|Q_i^{1/2}(a)|K^+\ra$, $i\ne 7,8$, are
quadratically divergent due to mixing with the operator
\beq
\cO_{sub}=(m_s+m_d)\bar s d-(m_s-m_d)\bar s\gamma_5d \ .
\eeq
To subtract this divergence, one uses the fact that the coefficients
of the parity even and odd components of $\cO_{sub}$ are fixed by
chiral and CPS symmetry \cite{Bernard:1985wf}. Then one constructs a
subtracted operator $Q_i-\alpha_i\cO_{sub}$ by imposing the constraint
\beq
\la 0|Q_i-\alpha_i\cO_{sub}|K^0\ra = 0\ ,
\label{eq:quadsub}
\eeq
to determine $\alpha_i$. Because CP-PACS works with degenerate quarks,
they obtain $\alpha_i$ from a derivative of \eq{eq:quadsub} with respect
to $m_s$ at the point $m_s=m_d$. RBC work with $m_s\ne m_d$ and obtain
$\alpha_i$ from the slope in $m_s-m_d$.

Note that this construction relies heavily on the fact that the
coefficients of $\bar s d$ and $\bar s\gamma_5d$ in $\cO_{sub}$ are
fixed.  In the absence of chiral symmetry, as when using Wilson
fermions, this is no longer the case and one finds that the matrix
elements discussed above are cubically divergent due to mixing with
$\bar s d$ and that the coefficient of this divergence cannot be
obtained by studying $K\to 0$ matrix elements. 

\medskip

Using leading order $\chi$PT, CP-PACS relate the $K\to\pi$ matrix
elements that they calculate to the desired $K\to\pi\pi$ matrix elements,
through:
\beq
\la Q_i\ra_I
 =C_If_\pi(m_K^2-m_{\pi}^2)\sum_k\frac{Z_{ik}}{Z_A^2}
\label{eq:qi0def}
\eeq
$$
\times\l[
\frac{\la\pi^+|Q_k^{\Delta I}-\alpha_k^{\Delta I}\cO_{sub}
|K^+\ra}{\la\pi^+|\bar u\gamma_0 \gamma_5d|0\ra
        \la 0|\bar s\gamma_0\gamma_5 u|K^+\ra}\r]_\chi {\rm [GeV^3]}
$$
\beq
\la Q_j\ra_I
 = -C_If_\pi\sum_k\frac{Z_{jk}}{Z_A^2}
\eeq
$$
\times
\l[m_P^2\,
\frac{\la\pi^+|Q_k^{\Delta I}|K^+\ra}{\la\pi^+|\bar u\gamma_0 \gamma_5d|0\ra
        \la 0|\bar s\gamma_0\gamma_5 u|K^+\ra}\r]_\chi {\rm [GeV^3]}
\ ,$$
with $i=1,2,\cdots,6,9,10$, $j=7,8$ and where $\Delta I=1/2(3/2)$
and $C_I=\sqrt{3/2}(\sqrt{3})$ for $I=0(2)$. The $Z_{ij}/Z_A^2$ are
the constants required to match the lattice results onto the
NDR-$\msbar$ scheme. $\alpha_i^{3/2}=0$ and $\alpha_i^{1/2}$ is
obtained as described around \eq{eq:quadsub}. $m_P$ is the mass of the
pseudoscalar meson in the simulation. $[\ldots]_\chi$ indicates that
the term in brackets is linearly extrapolated to the chiral limit.

Their results for the plaquette and RG-improved actions are very
similar.  We will therefore concentrate on the former, because they
can be compared directly to RBC's results, obtained with the same
parameters.

\begin{figure}[t]
\centerline{\epsfxsize=7.5cm\epsffile{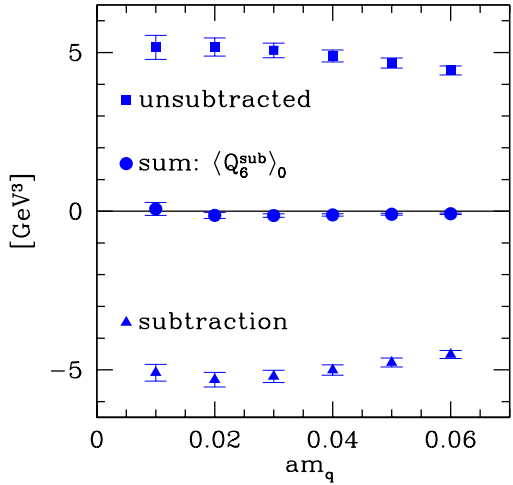}}
\vspace{-1.5cm}
\caption{\label{fig:q6sub}\small\it Subtraction of
the quadratic divergence which appears in the construction of $\la
Q_6\ra_0$ given in \eq{eq:qi0def}, for different
values of the degenerate quark mass, $m_q$.}
\vspace{-0.6cm}
\end{figure}
\fig{fig:q6sub} shows the subtraction of the power divergence in the
calculation of $\la Q_6\ra_0$ performed by CP-PACS, as a function of
their degenerate quark mass. The cancellation is severe and leads to a
value for $\la Q_6^{sub}\ra_0$~\footnote{By $Q_i^{sub}$, I mean the 
operator with power divergences subtracted, but that still
requires a logarithmic renormalization.} which is very roughly 40 times
smaller than the individual contributions. That a signal remains at
all must be due to the strong statistical correlation between the two
terms.

\medskip

While CP-PACS presented preliminary results for the matrix elements,
$\la Q_i(2\,\gev)\ra_{I}^{NDR}$, with $i=1,\cdots 10$, and $I=0,2$, it
is important to remember that this is a very difficult calculation and
that premature phenomenological consequences should not be drawn. I
have therefore chosen not to quote these results here. Let me instead
comment on the problems they may be confronted with.

One of the features of their results is that some of the $\alpha_s$
corrections in the matching of $\la Q_i^{sub}\ra_0$ onto the
NDR-$\msbar$ scheme are large, as large as 50\% in some instance. It
will be interesting to see what RBC finds with its non-perturbative
matching.

\begin{figure}[t]
\centerline{\epsfxsize=7.3cm\epsffile{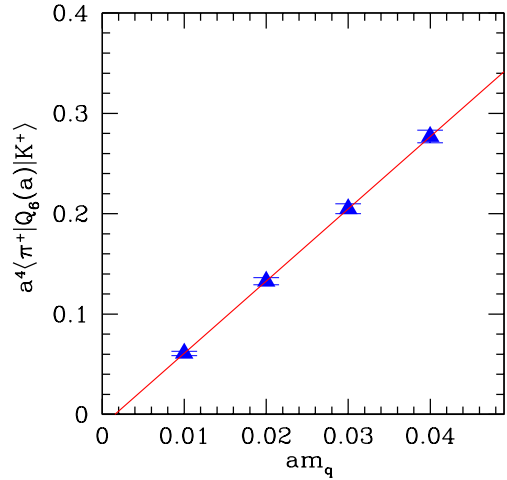}}
\vspace{-1.5cm}
\caption{\label{fig:q6unsubvsmq}\small\it $K^+\to\pi^+$
matrix element of the bare operator $Q_6$. The line corresponds to an
uncorrelated linear fit to the data.}
\vspace{-0.6cm}
\end{figure}

Also, as mentioned above, the fact that the divergence is quadratic in
the $\Delta I=1/2$ channel is a consequence of chiral
symmetry. Because chiral symmetry is only approximate with domain-wall
fermions, we actually expect there to be a cubic divergence due to
mixing with $\bar s d$, as in the Wilson fermion case. However,
instead of being of order 1, the coefficient of this divergence will
be exponentially small, leading to a term of the form $e^{-cN_5}\bar s
d/a^3$, which is formally of order $m_{res}\bar s d/a^2$. Such a term
will not be subtracted by the condition of \eq{eq:quadsub}. It will
appear as a non-vanishing intercept at $m_q=-m_{res}$ in a plot of the
matrix element versus quark mass of an operator such as $Q_6$, which
should vanish in the chiral limit. Such an intercept is seen by RBC,
as shown in \fig{fig:q6unsubvsmq}. 
This intercept is still reasonably small on the scale of the
unsubtracted $\la\pi^+|Q_6(a)|K^+\ra$ matrix element of
\fig{fig:q6unsubvsmq}, at the values of $am_q$ at which the calculation
is performed. However, it is large when converted to the scale of $\la
Q_6^{sub}\ra_0$ in \fig{fig:q6sub} and increases like $1/m_P^2$
when $m_P$ is decreased, as the definition of \eq{eq:qi0def}
implies.~\footnote{Here I am assuming that the intercept measured by
RBC is similar to the one CP-PACS would measure: the simulation
parameters are identical.} Of course, the estimate of the size of this
unphysical contribution will depend sensitively on how the chiral
extrapolation of $\la\pi^+|Q_6(a)|K^+\ra$ is performed, a delicate
question in quenched QCD and in a finite volume.~\footnote{Note that
the size of the intercept is not inconsistent with it being of order
$-2m_{res}\la\bar qq\ra/(a^2f_\pi^2)$, which is the chiral value
of $m_{res}\la\pi^+|\bar s d|K^+\ra/a^2$.} But this should be taken as
a warning that such effects may be important and should not be
ignored. One way to subtract such an unphysical intercept is to obtain the
chiral-limit matrix element from the slope in the dependence of $\la
\pi^+|Q_6^{sub}|K^+\ra$ on $m_q$. This is the procedure advocated by
RBC \cite{bob} and I would favor it over the use of
\eq{eq:qi0def}.

It should also be remembered that what is calculated are $K\to\pi$
matrix elements where the $K$ and the $\pi$ are degenerate and at
rest. These matrix elements, suitably normalized, are then
extrapolated to the chiral limit and translated into $K\to\pi\pi$
matrix elements using leading order $\chi$PT. This procedure thus
requires a good control over chiral extrapolations, which is not
trivial in the quenched approximation and in a finite
volume.~\footnote{This issue is currently being investigated by the
RBC collaboration \cite{bob}.} Furthermore, the leading order $\chi$PT
relations between $K\to\pi$ and $K\to\pi\pi$ neglect effects which may
be important, such as final-state interactions.

Finally, results involving an active charm quark have not yet been
presented. Given the fact that the charm is not that heavy, its
contribution as an active quark may be important.

\section{$K\to\pi\pi$ from unphysical $K\to\pi\pi$}
\label{sec:kpipi}

This approach is complementary to the $K\to\pi$ and $K\to 0$ method
discussed above. Here, four-point functions are used to compute the
matrix elements for $K\to\pi\pi$ transitions at unphysical values of
the mesons' masses and momenta, imposed by computational and theoretical
limitations. Then, low order $\chi$PT
is used to extrapolate the result to the physical point. While this
method is numerically more demanding because of the four-point
functions required, renormalization is simplified. In fact, this
approach appears to be the only one possible when considering $\Delta
I=1/2$ transitions in the absence of the GIM mechanism (as is
the case when studying $\epsilon'$) with fermions which explicitly
break chiral symmetry, such as Wilson fermions.

The SPQcdR collaboration are undertaking a quenched study of the
$\Delta I=1/2$ rule and of $\epsilon'$ using non-perturbatively
$\ord{a}$-improved Wilson fermions \cite{guido} and NPR matching \`a
la \cite{Martinelli:1995ty}. All mesons are taken at
rest. They work
with degenerate quarks, i.e. $m_u=m_d=m_s$, but consider also the
situation where the quarks in the kaon and in the pions
have masses tuned so that $m_K=2m_\pi$. With pions at rest, this
latter situation yields $K\to\pi\pi$ at threshold.

In the $\Delta I=3/2$ channel, renormalization is particularly simple
as there is no mixing with lower-dimension operators and CPS symmetry
guarantees that there is no mixing with wrong-chirality operators of
dimension six \cite{Bernard:1988pr}. They presented an exploratory
study of the matrix element $\la Q_2+Q_1\ra_2$, which constitutes the
denominator of the $\Delta I=1/2$ rule. This amplitude has already
been studied in the quenched approximation with unimproved Wilson
fermion, perturbative matching and degenerate quarks
\cite{Gavela:1988bd,Bernard:1989zj,Aoki:1998ev}, as reviewed last year
\cite{Kuramashi:2000gt}. By accounting for finite-volume effects and
chiral logarithms \cite{Golterman:1997wb}, and using modern techniques
in perturbative renormalization, the authors of the most recent study
\cite{Aoki:1998ev} were able to reconcile, with experiment, the
physical amplitude obtained from the chiral-limit lattice results. The
pioneering studies of \cite{Gavela:1988bd,Bernard:1989zj} had found a
discrepancy by a factor of roughly two. Non-perturbative matching and
$\ord{a}$-improvement, as well as the situation where $m_K=2m_\pi$, as
considered by SPQcdR, should help further clarify the situation for
these decays.

SPQcdR also presented exploratory results from the first $K\to\pi\pi$
calculation of the amplitudes $\la Q_{7,8}\ra_2$
\cite{guido}. Combining their results with those obtained from
$K\to\pi$ amplitudes should help obtain more reliable predictions for
these matrix elements.

\medskip

In the $\Delta I=1/2$ channel, SPQcdR are considering mixing with the
lower dimension operators~\footnote{$\tilde\cO_G$ is actually
subleading in the chiral expansion \cite{Deshpande:1994vp}.}
\bea
\cO_P&=&(m_s-m_d)\bar s\gamma_5 d\\
\tilde\cO_G&=&(m_s-m_d)\bar s\sigma_{\mu\nu}\tilde G_{\mu\nu} d\nn
\ ,\eea
where the factor of $(m_s-m_d)$ is required by CPS symmetry. (The
$\Delta S=1$ operators of \eq{eq:ds1ops} are all even under CPS.)
Clearly, working with $m_u=m_d=m_s$ eliminates this mixing completely
\cite{Bernard:1988pr}.  It is shown in \cite{Dawson:1998ic} that the
choice $m_K=2m_\pi$ with all mesons at rest also simplifies the
renormalization because no momentum is transfered by the $\Delta S=1$
operator. SPQcdR is pursuing both these avenues \cite{guido}.

\section{Non-leptonic weak decays from finite-volume correlation functions}

\label{sec:finitev}

Determinations of $K\to\pi\pi$ amplitudes, both from $K\to\pi$ (and
$K\to 0$) matrix elements and from unphysical $K\to\pi\pi$ amplitudes,
rely on low order $\chi$PT. They therefore neglect chiral corrections
which may be important. Aside from this limitation, it would be
satisfying to be able to calculate $K\to\pi\pi$ amplitudes in full,
directly on the lattice. Can this be done?

\subsection{Euclidean correlation functions and the ``Maiani-Testa no-go
theorem''}

The lattice method provides approximate estimates of
euclidean correlation functions. While the Osterwalder-Schrader
theorem guarantees that these correlation functions can, in principle,
be analytically continued back to Minkowski space, such a procedure is
unstable in practice, when applied to approximate data. This led
Maiani and Testa to investigate what can be extracted from euclidean
correlation functions without analytic continuation \cite{Maiani:1990ca}.

They considered a correlation function which, in the case of $K\to\pi\pi$
decays, can be written
\beq
C_{K\pi\pi}(t_1,t_2,t_3;\vec{p},-\vec{p})\equiv 
\label{eq:mtecf}
\eeq
$$
\la \pi_{\vec{p}}(t_1)
\pi_{-\vec{p}}(t_2)\cH_{\Delta S=1}(0) K_{\vec{0}}(-t_3)\ra
\ ,
$$
where
\beq
\pi_{\vec{p}}(t)\equiv \int d^3x\,e^{-i\vec{p}\cdot\vec{x}}\pi(\vec{x},t)
\ ,
\eeq
and similarly for $K_{\vec{p}}(t)$. Then they investigated the
behavior of this correlation function as a function of $t_2>0$ in the
limit $t_1,\,t_3\to +\infty$. 

So as to disentangle euclidean from finite-volume effects, they chose
to work in the limit of asymptotically large volumes. This has for
consequence that their spectrum of two-pion final states is continuous
which, in turn, means that when $t_2$ is taken to $+\infty$, only the
ground state contribution can be picked out. Thus, they found that the
euclidean correlation function of \eq{eq:mtecf} only gave them the
matrix element $\la\pi\pi|\cH_{\Delta S=1}|K\ra$ with all mesons at
rest, which is certainly not the physical kinematics. The further
found that $1/\sqrt{t_2}$ corrections to the $t_2\to +\infty$ result
contained information about the $\pi\pi$ scattering length.

This direction has been pursued by Ciuchini {\it et al.}\ who argue
that they can reconstruct the desired weak decay amplitudes under the
assumption that final-state interactions are dominated by nearby
resonances and that the couplings to these resonances is smooth in the
external momenta \cite{Ciuchini:1996mq}. Lin {\it et al.}\ are
currently working on eliminating the need for these assumptions
\cite{linetal}.

\subsection{Two-pion states in finite volume}

In \cite{Lellouch:2000pv}, a different route has been followed. In
present day simulations, lattice have sides, $L$, of order a few
fermi. This means that the spectrum of two-pion states on such
lattices is discrete: the quantum of momentum is $\Delta p=
2\pi/L=1.2\,\gev/L[\fm]$. Thus, this spectrum is far from
continuous.  This statement can, in fact, be made much more precise.

Consider a box of volume $L\times L\times L$ with periodic boundary
conditions and sides $L\ge 3\,\fm$.~\footnote{This guarantees that
single-particle states resemble their infinite-volume counterparts up
to terms exponentially small in $L$.} In such a box, it was shown by
L\"uscher \cite{Luscher:1991ux} that two-pion energies, in the $A_1^+$
(``spin-0'') sector and center of mass frame, and below the four-pion
threshold, are given by
\beq
W_\ell=2\sqrt{m_\pi^2+k_\ell^2}, \quad\quad \ell=1,2,\ldots
\label{eq:2pien}
\eeq
$$
\ell\pi-\delta_0^I(k_\ell)=\phi(q_\ell), \quad\quad 
q_\ell=\frac{k_\ell L}{2\pi}
$$
where $\delta_0^I(k)$ is the $\pi\pi$ scattering phase shift in the
spin-0 and isospin-$I$ channel and $\phi(q)$ is a known kinematical
function. 

To get some idea of what these equations say, one can expand their
solution in inverse powers of $L$ and one finds ($\ell\le 6$)

\beq
W_\ell=2\sqrt{m_\pi^2+\ell(2\pi/L)^2}+\ord{1/L^3}
\ .\eeq
This is the energy of two free pions, each with momentum
$\sqrt{\ell}(2\pi/L)$, up to corrections which go like one over the
volume and which are determined by the phase shift.

\begin{figure}[t]
\centerline{\epsfxsize=8cm\epsffile{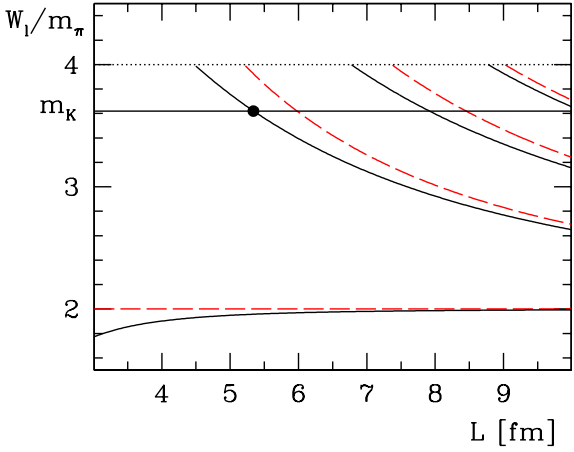}}
\vspace{-1cm}
\caption{\label{fig:2pien}\small\it Two-pion energy
spectrum in QCD below the inelastic threshold and as a function of the
length of the box's sides. The dashed curves correspond to free
pions. The solid curves correspond to interacting pions in the
$I=J=0$ channel. Also shown is the kaon mass, $m_K$, and the
intersection of the first excited state ($\ell=1$) with this mass
(dot).}
\vspace{-0.6cm}
\end{figure}

In \fig{fig:2pien}, the two-pion energies are plotted in units of the
pion mass as a function of the length, $L$, of the box's sides. The
dashed curves are the energies of two free pions. The solid curves are
the energies of two interacting pions with $I=J=0$, as derived from
\eq{eq:2pien}, with $\delta_0^0$ calculated at one-loop in $\chi$PT
\cite{Gasser:1983kx,Gasser:1991ku} and taken from
\cite{Knecht:1995tr}. Please note that $\chi$PT is only used here for
the purpose of illustrating how \eq{eq:2pien} works. It is not in any
way required by the method.

The first thing to note is that the spectrum is far from
continuous for any value of $L$ one may hope to reach in numerical
simulation. The second point is that the corrections to the spectrum
brought about by $\pi\pi$ interactions are not that large. This is
due to the fact that they appear suppressed by $1/L^3$
and the volumes considered are not that small.  A
third point is that \eq{eq:2pien} can be
turned around. Thus, a numerical study of the finite-volume two-pion
spectrum can be converted into a determination of the phase shift
$\delta_0^I(k)$ \cite{Luscher:1991ux}.  

\subsection{$K\to\pi\pi$ decays in finite volume}

Having established that the two-pion levels are discrete, one can then
imagine isolating $K$ decays to some of these excited two-pion levels,
thus determining the matrix element~\footnote{A possible approach to
isolating excited two-particle levels is discussed in
\cite{Luscher:1990ck}.}
\beq
 M^I_\ell\equiv
\la(\pi\pi)_I\ell|\cH_{\Delta S=1}|K\ra 
\label{eq:fvamp}
\ .\eeq
In \eq{eq:fvamp}, all finite volume states are normalized to one: $\la
K|K\ra=\la(\pi\pi)_I\ell|(\pi\pi)_I\ell\ra=1$.

Furthermore, having obtained $\delta_0^I(k)$ from a numerical study of
the finite-volume two-pion spectrum, one can imagine tuning $L$ in
such a way that the $n^{th}$ $\pi\pi$ state has energy
$W_n=m_K$. This is illustrated by a dot in \fig{fig:2pien}, for
$n=1$. The corresponding transition matrix element, $M^I_n$, conserves
both energy and momentum and the kaons and pions have their physical
masses. It is not, however, a finite-volume matrix element that we are
after. What we want is the matrix element which describes the decay of
a kaon at rest into a state of two asymptotic pions with energy $m_K$,
i.e. $T^I_n\equiv\la(\pi\pi\, out)_I;m_K|\cH_{\Delta S=1}|K\ra$. The hard work
enters in showing that these two matrix elements are related.

In \cite{Lellouch:2000pv}, it is shown that the infinite-volume
amplitude, $T^I_n$, is related to the corresponding finite-volume
matrix element, $M^I_n$, through~\footnote{In the study of the
normalization of $\pi\pi$ wavefunctions in finite and infinite volume,
as well as in the perturbative check, both of which are discussed
below, a relation which holds for $W_n\ne m_K$ can
also be obtained. It is identical to \eq{eq:merel}, except that the factor of
$\l(m_K/k_n\r)^3$ is replaced by $W_n^2 m_K/k_n^3
$. Whether this result holds more generally has yet to be verified.}
\beq
\left|T^I_n\right|^2=8\pi\, L^6\left\{q\,\phi'(q)+
  k\,\delta_0^{I\prime}(k)\right\}_{k=k_n}
\label{eq:merel} 
\eeq
$$
\times  \l(\frac{m_K}{k_n}\r)^3
\left|M^I_n\right|^2
\ .
$$
This result assumes that $W_n<4 m_\pi$ and that the final two-pion
state $|(\pi\pi)_In\ra$ is not degenerate (i.e. $n<8$). It is
accurate up to exponentially small corrections in $L$.

There are various ways to proceed to prove this relation
\cite{Lellouch:2000pv}.  One approach is to study the normalization of
$\pi\pi$ wavefunctions in finite and infinite volume, with the
quantum-mechanical formalism of
\cite{Luscher:1991ux,Luscher:1986dn}. Then one uses the fact that the
transition matrix elements probe the $S$-wave component of the two-pion
wavefunction near the origin and that this component differs in finite
and infinite volume only in its normalization and possibly phase.

Instead, one can switch on the weak
interaction ($\cH_{\Delta S=1}$) and compute its influence on $\pi\pi$
energies. This can be done directly, using ordinary quantum-mechanical
perturbation theory, or one may start from \eq{eq:2pien}, taking into
account the effect of the weak interaction on the scattering
phase. A comparison of the two results yields the relation of
\eq{eq:merel}.

One can also verify this relation by working out both finite and
infinite-volume amplitudes in perturbation theory, in a low-energy
effective theory such as $\chi$PT. Because \eq{eq:merel} does not
depend on the details of the dynamics of the system, a highly
simplified model was chosen in \cite{Lellouch:2000pv}, so as not to
obscur the relation between the amplitudes, with complicated
interactions and unnecessary quantum numbers. Since this calculation
does not rely on \eq{eq:2pien}, it can provide additional confidence
in the correctness of \eq{eq:merel} and illustrates how this relation
plays out in the context of a relativistic field theory.

Recently, yet another derivation of \eq{eq:merel} was proposed
\cite{linetal,Testa:2000ce}. The relation between this new approach
and that of \cite{Lellouch:2000pv} should be clarified soon
\cite{linetal}.

\subsection{Application to the $\Delta I=1/2$ rule}

Let $n=1$ with $W_n=m_K$. One possible statement of the $\Delta
I=1/2$ rule is that $|T^0_n|/|T^2_n|\simeq 22$. What does this rule
look like in finite volume?

For the purpose of illustration, we suppose that the scattering phases
$\delta^I_0$ are accurately given by their one-loop expression in
$\chi$PT \cite{Knecht:1995tr}.~\footnote{Again, when the approach we
have laid out is carried out in full, these phases will be obtained
from a numerical study of the two-pion energy spectrum in finite
volume.} Then, using the two-pion energy formulae of \eq{eq:2pien},
the size of the box, $L$, required for the first excited state ($n=1$)
to have energy $W_1$ equal to the kaon mass can be calculated in the
two isospin channels. The results are shown in \tab{tab:mereleg}. Once
the size of the boxes is fixed, it is straightforward to obtain the
proportionality factor which appears in \eq{eq:merel}. The different
contributions to this factor are also given in \tab{tab:mereleg}.
\begin{table}
\begin{center}
\caption{\label{tab:mereleg}\small\it Size of the box, $L$, required
for the first excited, two-pion state to have energy equal to $m_K$ in
the isopin 0 and 2 channels. Also given are the contributions to the
proportionality factor which appears in \eq{eq:merel}.}
\begin{tabular}{ccccc}
\\
\hline
\hline\\[-4mm]
$I$ & $L\,[\fm]$ & $q$ & $q\,\phi'(q)$ 
& $k\,\delta_0^{I\prime}(k)$ \\
\hline
0 & 5.34 & 0.89 & 4.70 & 1.12\\
2 & 6.09 & 1.02 & 6.93 & $-0.09$\\
\hline
\hline
\end{tabular}
\end{center}
\vspace{-0.6cm}
\end{table}
One ends up with
\beq
|T^0_n|=44.9\times|M^0_n|,\qquad |T^2_n|=48.7\times 
|M^2_n|,
\label{eq:mereleg}
\eeq
and
\beq
\l|T^0_n/T^2_n\r|=0.92\times \l|M^0_n/M^2_n\r|
\ .
\label{eq:merelegratio}
\eeq
While the factors in \eq{eq:mereleg} look large, they are mainly due
to the relative normalization of free states in finite and infinite
volume.  Indeed, in the free case (i.e. with the scattering phases set
to zero), the relation is $|T^I_n|=47.7\times|M^I_n|$. Thus, the
effect of interactions is relatively small despite the large
difference between the scattering phases in the two isospin channels
(approximatively $45^o$ for $k=\sqrt{m_k^2/4-m_\pi^2}$).  This means,
in particular, that if QCD is to reproduce the $\Delta I=1/2$
enhancement, the large factor will have to come from the ratio of
finite volume matrix elements, $\l|M^0_n/M^2_n\r|$. In fact, as
\eq{eq:merelegratio} suggests, the effect should even be slightly
larger in finite volume.

\subsection{Summary}

$K\to\pi\pi$ rates can, in principle, be obtained from the lattice
without any model assumptions and without analytic continuation
\cite{Lellouch:2000pv}.  This requires working on lattices whose
sides, $L$, are greater than $5\,\fm$. On these lattices, the effect
of interactions on the proportionality factor relating the transition
matrix elements in finite and infinite volume was found to be
small. This means that the finite-volume amplitudes must incorporate
most of the physics which enters the determination of their infinite
volume counterparts. In particular, if QCD is to reproduce the $\Delta
I=1/2$ enhancement, this enhancement will be clearly visible in finite
volume.

For the approach to be fully self-contained, the strong scattering
phases, which are required to relate the finite and infinite volume
amplitudes, should also be determined on the lattice. As shown by
L\"uscher many years ago \cite{Luscher:1991ux,Luscher:1986dn} and as
briefly discussed above, this can again be done using finite-volume
techniques. Here, the recent high-precision determination of $\pi\pi$,
$S$-wave scattering lengths by Colangelo {\it et al.}\
\cite{Colangelo:2000jc} should provide a good test of some of the
lattice techniques required to undertake such studies.

The same ideas as those described above can be applied to baryon
decays, such as $\Lambda\to N\pi$, $\Sigma\to N\pi$ and $\Xi\to
\Lambda\pi$, as well as to any other decay in which the final-state
particles scatter only elastically.

There are, however, many potential hurdles in implementing the
approach.  It may be difficult, for instance, to extract excited
$\pi\pi$ states in practice. Furthermore, since the unitarity of the
underlying theory played an essential r\^ole in the argument leading
to the relation of \eq{eq:merel}, it is not clear to what extent this
relation holds in quenched QCD. Quenched $\chi$PT, along the lines of
\cite{Bernard:1996ez,Golterman:1997wb}, should help shed light on this
problem. Finally, the approach taken here only applies to situations
where the particles in the final state scatter elastically. It says
very little about what should be done when they can scatter
inelastically, as for example in $B\to\pi\pi$, $D\to K\pi$, $\ldots$
decays.

\section{Conclusions}

\label{sec:ccl}

This has been an exciting year. New ideas for dealing with
non-leptonic weak decays and for reducing operator mixing with Wilson
fermions have been proposed. 

There are new chiral perturbation theory results relevant for
extracting $K\to\pi\pi$ amplitudes from $K\to\pi$ and $K\to 0$ matrix
elements and from unphysical $K\to\pi\pi$ matrix elements.

There are also new numerical results, many of which are still preliminary,
for:
\vspace{-0.15cm}
\begin{itemize}
\item $\Delta S=2$ matrix elements relevant for $K^0$-$\bar K^0$
mixing in the standard model and beyond, with domain-wall and
$\ord{a}$-improved Wilson fermions;
\vspace{-0.2cm}
\item $K\to(\pi\pi)_I$ decays in the CP conserving and violating
sectors, obtained from $K\to\pi$ and $K\to 0$ matrix elements, with
domain-wall (for $I=0,2$) and $\ord{a}$-improved Wilson (for $I=2$)
fermions.
\end{itemize}

One thus expects, in the near future, lattice determinations of
$\re\epsilon'/\epsilon$ and studies of the $\Delta I=1/2$-rule,
hopefully with an active charm. In interpreting these results, when
they come out, it will be important to keep in mind the difficulties
of this calculation and the limitations of the $K\to\pi$ and $K\to 0$
approach, as discussed in \sec{sec:kpipifromdwf}.

There should also soon be results for $K\to(\pi\pi)_{I=2}$ decays in
the CP conserving and violating sectors, from $K\to\pi\pi$ matrix
elements, obtained with $\ord{a}$-improved Wilson
fermions. Kinematical situations other than the traditional
$m_K=m_\pi$ with the two pions at rest will be investigated. It is
further expected that these studies will be extended to $\Delta I=1/2$
transitions.

In any event, the coming year may yet be even more exciting.

\medskip
\noindent{\it Acknowledgements}

I would like to thank S.\ Aoki, T.\ Blum, N.\ Christ, M.\ Golterman,
M.\ Knecht, C.-J.\ D.\ Lin, M.\ L\"uscher, G.\ Martinelli, R.\
Mawhinney, J.-I.\ Noaki, C.\ Sachrajda, A.\ Soni, Y.\ Taniguchi and
M.\ Testa for sharing their results with me and/or for interesting
discussions. I am grateful to David Lin for his careful reading of the
manuscript. Many thanks to the organizers, also, for a delightful
conference. Work supported in part by TMR, EC-Contract
No.~ERBFMRX-CT980169.


\end{document}